\documentclass[prl,letterpaper,english,reprint,aps,superscriptaddress]{revtex4-2}

\usepackage{babel,calc,amsmath,amsthm,amssymb,graphicx,subfigure,xcolor}
\usepackage{mathdots}
\usepackage[T1]{fontenc}
\usepackage[unicode=true]{hyperref}
\usepackage{booktabs}
\usepackage{mathtools}
\usepackage{multirow}
\usepackage{lipsum}
\usepackage[normalem]{ulem}
\definecolor{prllinkblue}{rgb}{0.1804, 0.1882, 0.5726}
\hypersetup{
     colorlinks=true,       		
     linkcolor=prllinkblue,          	
     citecolor=prllinkblue,             
     urlcolor=prllinkblue,          
 }
 
\newtheorem*{theorem*}{Theorem}

\newtheorem*{corollary*}{Corollary}

\newtheorem*{lemma*}{Lemma}

\newtheorem*{proposition*}{Proposition}
\theoremstyle{definition}

\newtheorem*{definition*}{Definition}
\theoremstyle{remark}

\newtheorem*{remark*}{Remark}

\newcommand{\ket}[1]{\left\vert#1\right\rangle}
\newcommand{\bra}[1]{\left\langle#1\right\vert}
\newcommand{\braket}[1]{\left\langle#1\right\rangle}

\newif\ifdebug

\debugtrue

\ifdebug
\definecolor{zhliu}{rgb}{0.60, 0.12, 0.24}

\newcommand{\note}[1]{{\color{orange}{#1}}}
\newcommand\delete{\bgroup\markoverwith{\textcolor{zhliu}{\rule[0.5ex]{2pt}{0.8pt}}}\ULon}

\else

\newcommand{\note}[1]{\ignorespaces}
\newcommand{\delete}[1]{\ignorespaces}

\fi

\begin{document}
\renewcommand{\figurename}{Fig.}

\title{Experimental quantum phase discrimination enhanced by controllable indistinguishability-based coherence}

\author{Kai~Sun}
\author{Zheng-Hao~Liu}
\author{Yan~Wang}
\author{Ze-Yan~Hao}
\author{Xiao-Ye~Xu}

\author{Jin-Shi~Xu}
\email{jsxu@ustc.edu.cn}

\author{Chuan-Feng~Li}
\email{cfli@ustc.edu.cn}

\author{Guang-Can~Guo}

\affiliation{CAS Key Laboratory of Quantum Information, University of Science and Technology of China, Hefei 230026, People's Republic of China}
\affiliation{CAS Centre For Excellence in Quantum Information and Quantum Physics, University of Science and Technology of China, Hefei 230026, People's Republic of China}

\author{Alessia Castellini}
\affiliation{Dipartimento di Fisica e Chimica - Emilio Segr\`e, Universit\`a di Palermo, via Archirafi 36, 90123 Palermo, Italy}

\author{Ludovico Lami}
\affiliation{Institut f\"ur Theoretische Physik und IQST, Universit\"at Ulm,
Albert-Einstein-Allee 11, D-89069 Ulm, Germany}

\author{Andreas Winter}
\affiliation{ICREA \& F\'{i}sica Te\'{o}rica: Informaci\'{o} i Fen\'{o}mens Qu\`{a}ntics, Departament de F\'{i}sica, Universitat Aut\`{o}noma de Barcelona, ES-08193 Bellaterra
(Barcelona), Spain}

\author{Gerardo Adesso}
\affiliation{School of Mathematical Sciences and Centre for the Mathematics and Theoretical Physics of Quantum Non-Equilibrium Systems, University of Nottingham, University Park, Nottingham NG/ 2RD, United Kingdom}

\author{Giuseppe Compagno}
\affiliation{Dipartimento di Fisica e Chimica - Emilio Segr\`e, Universit\`a di Palermo, via Archirafi 36, 90123 Palermo, Italy}

\author{Rosario Lo Franco}
\email{rosario.lofranco@unipa.it}
\affiliation{Dipartimento di Ingegneria, Universit\`{a} di Palermo, Viale delle Scienze, Edificio 6, 90128 Palermo, Italy}

\date{\today}

\begin{abstract}

 Quantum coherence, a basic feature of quantum mechanics residing in superpositions of quantum states, is a resource for quantum information processing. Coherence emerges in a fundamentally different way for nonidentical and identical particles, in that for the latter a unique contribution exists linked to indistinguishability which cannot occur for nonidentical particles. We experimentally demonstrate by an optical setup this additional contribution to quantum coherence, showing that its amount directly depends on the degree of indistinguishability and exploiting it to run a quantum phase discrimination protocol.
Furthermore, the designed setup allows for simulating Fermionic particles with photons, thus assessing the role of particle statistics (Bosons or Fermions) in coherence generation and utilization. 
Our experiment proves that independent indistinguishable particles can supply a controllable resource of coherence for quantum metrology.
  
  
\end{abstract}

\maketitle

\textit{Introduction.}---A quantum system can reside in coherent superpositions of states, giving rise to  nonclassicality \cite{gleason1957, ks1967} which implies the intrinsic probabilistic nature of predictions in the quantum realm \cite{schrodinger, wigner, renner2018, wiseman2020,Bera_2017}. Besides this fundamental role, quantum coherence is also at the basis of quantum algorithms \cite{shor1994, obrien2012, cleve1998, procopio2015, aaronson2011, google2019} and, from the modern information-theoretic perspective, constitutes a paradigmatic basis-dependent quantum resource \cite{andreas2016, adesso2017, chitambar2019}, providing a quantifiable advantage in certain quantum information protocols.

For a single quantum particle, coherence emerges when the particle is found in a superposition of states in a given basis of the Hilbert space. For multiparticle compound systems, the physics underlying the emergence of coherence is richer and strictly connected to the nature of the particles, with fundamental differences for nonidentical and identical particles. In fact, states of identical particle systems can manifest coherence even when \textit{no} particle resides in superposition states, provided that the wavefunctions of the particles overlap \cite{saro2019pra,Chin2019,walmsleyPRA2017}. In general, a special contribution to quantum coherence arises thanks to the spatial indistinguishability of identical particles which cannot exist for nonidentical (or distinguishable) particles \cite{saro2019pra}. 
Recently, it has been found that the aptitude of spatial indistinguishability of identical particles can be exploited for entanglement generation \cite{saro2018prl}, applicable even for spacelike-separated quanta \cite{castellini2019pra} and against preparation and dynamical noises \cite{saro2020npj, NosratiPRA, alexander2018}. 
The presence of entanglement is a signature that the bipartite system as a whole carries coherence even when the individual particles do not, the amount of this coherence being dependent on the degree of indistinguishability. 
We name this specific contribution to quantumness of compound systems as ``indistinguishability-based coherence'', as a difference with the more familiar ``single-particle superposition-based coherence''. 
Indistinguishability-based coherence qualifies in principle as an exploitable resource for quantum metrology \cite{saro2019pra}. However, it requires sophisticated control techniques to be harnessed, especially in view of its nonlocal nature. Moreover, a crucial property of identical particles is the exchange statistics, while operating both Bosons and Fermions in the same setup is generally challenging.

\begin{figure}[t]
  \centering
  \includegraphics[width=0.48\textwidth]{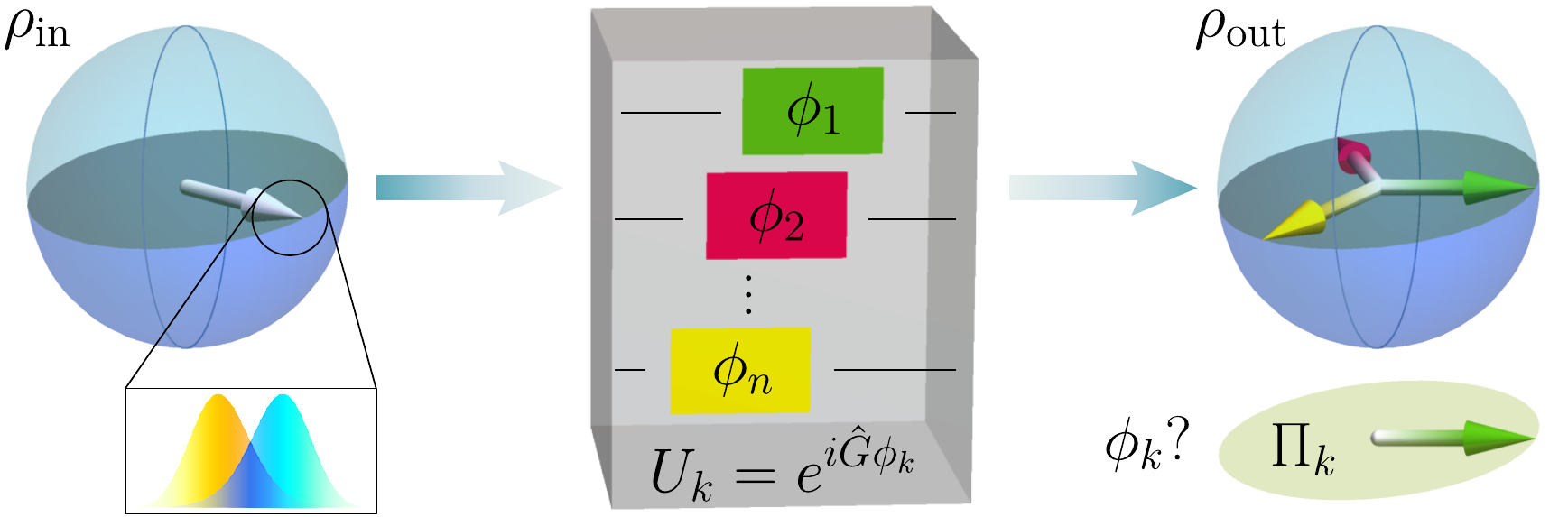}
  \caption{Illustration of the indistinguishability-activated phase discrimination task. A resource state $\rho_\mathrm{in}$ that contains coherence on a computational basis is distilled from spatial indistinguishability. The state then enters a black box which implements a phase unitary $\hat{U}_k=e^{i\hat{G}\phi_k}, k\in\{1, \ldots, n\}$ on $\rho_\mathrm{in}$. The goal is to determine the $\phi_k$ actually applied through the output state $\rho_\mathrm{out}$: indistinguishability-based coherence provides operational advantage to the task.}
  \label{fig:theory}
\end{figure}

In this work, we experimentally investigate the operational contribution to quantum coherence stemming from spatial indistinguishability of identical particles. By virtue of our recently developed photonic architecture capable of tuning the indistinguishability of two uncorrelated photons \cite{sun2020}, we observe the direct connection between degree of indistinguishability and amount of coherence, and show that indistinguishability-based coherence can be concurrent with single-particle superposition-based coherence. In particular, we demonstrate that it has operational implications, providing a quantifiable advantage in a phase discrimination task \cite{napoli2016, ringbauer2018}, as depicted in Fig.~\ref{fig:theory}. Furthermore, we design a setup capable to test the impact of particle statistics in coherence production and phase discrimination for both Bosons and Fermions. This is accomplished by compensating for the exchange phase during state preparation, simulating Fermionic states with photons, which leads to statistics-dependent efficiency of the quantum task.



\textit{Indistinguishability-based coherence}---To formally recall the idea of coherence activated by spatial indistinguishability \cite{saro2019pra}, we first consider the basic scenario where the wavefunctions of two identical particles with orthogonal pseudospins, $\downarrow$ and $\uparrow$ overlap at two spatially-separated sites, L and R. Omitting the unphysical labeling of identical particles \cite{saro2016}, the state is described as $\ket{\Psi}=\ket{\psi\downarrow, \psi'\uparrow}$, with $\ket{\psi}=l\ket{\rm L}+r\ket{\rm R}$ and $\ket{\psi'}=l'\ket{\rm L}+r'\ket{\rm R}$ denoting the spatial wavefunctions corresponding to the two pseudospins. 
Let us use spatially localized operations and classical communication, i.e., the sLOCC-framework  \cite{saro2018prl}, to activate and exploit the operational coherence. Projecting onto the operational subspace $\mathcal{B}=\{\ket{L\sigma,R\tau}; \sigma,\tau=\downarrow,\uparrow\}$ yields the normalized conditional state \cite{saro2019pra}
\begin{align}\label{eq:psi}
  \ket{\Psi_\mathrm{LR}}=\frac{1}{\mathcal{N}_\mathrm{LR}^\Psi}(lr'\ket{\rm L\downarrow,R\uparrow}+\eta l'r\ket{\rm L\uparrow,R\downarrow}),
\end{align}
with $\mathcal{N}_\mathrm{LR}^{\Psi}=\sqrt{|lr'|^2+|l'r|^2}$, and the exchange phase factor $\eta=1(-1)$ originates from the Bosonic (Fermionic) nature of the indistinguishable particles. 
We see that, although each particle starts from an incoherent state (namely, $\ket{\psi\downarrow}$, $\ket{\psi'\uparrow}$) in the pseudospin computational basis, the final state $\ket{\Psi_\mathrm{LR}}$ overall resembles a coherent, nonlocally-encoded qubit state in the compound basis $\mathcal{B}$ under spatially local operations and classical communication (sLOCC). Also, considering that this coherence vanishes when the two particles are nonidentical thus individually addressable \cite{saro2019pra}, the emergence of coherence in $\ket{\Psi_\mathrm{LR}}$ essentially hinges on the spatial indistinguishability of the identical particles, in strict analogy to the emergence of entanglement between pseudospins \cite{saro2018prl,sun2020,Barros:20}.

The coherence of the state of Eq.~(\ref{eq:psi}) is independent of the Bosonic or Fermionic nature of the particles because of the specific choice of the initial single-particle states. However, in general, particle statistics plays a role in determining the allowed spatial overlap properties of identical particles and is thus crucial for the coherence of the overall state of the system. Hence, we shall extend our experimental investigation to a state where these fundamental aspects can be observed. Taking again a scenario with two indistinguishable particles, one of the particles is now initialized with innate coherence in the pseudospin basis, i.e., the initial two-particle state reads $\ket{\Psi}=\ket{\psi\downarrow,\psi's'}$, where $\ket{s'}=a\ket{\uparrow}+b\ket{\downarrow}$ with $|a|^2+|b|^{2}=1$. Projecting onto $\mathcal{B}$ generates the three-level distributed state \cite{saro2019pra}
\begin{align}
  \ket{\Phi_\mathrm{LR}} = \frac{1}{\mathcal{N}_\mathrm{LR}^\Phi} (&a l r' \ket{\rm L\downarrow, R\uparrow} + b (l r'+ \eta l' r) \ket{\rm L\downarrow, R\downarrow} \nonumber\\
  +&a \eta l' r \ket{\rm L\uparrow, R\downarrow}),
  \label{eq:phi}
\end{align} 
where $\mathcal{N}_\mathrm{LR}^\Phi={\sqrt{a^2(|l r'|^2+|l' r|^2)+b^2|l r'+ \eta l' r|^2}}$. In this state, indistinguishability-based coherence coexists with single-particle superposition-based coherence, giving rise to an overall multilevel coherence in the operational basis $\mathcal{B}$. 

\begin{figure*}[t!]
  \centering
  \includegraphics[width=0.7\textwidth]{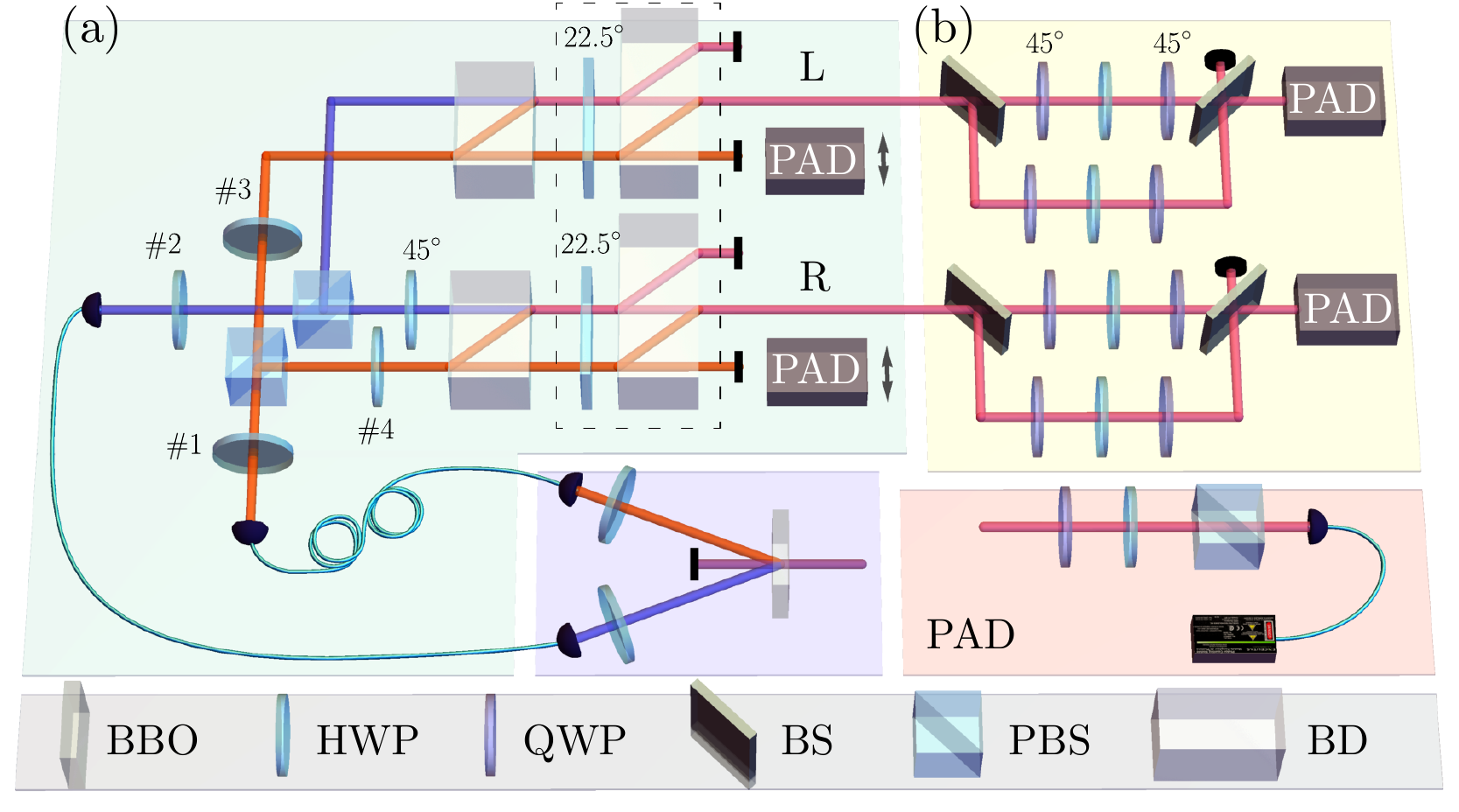}
  \caption{Experimental configuration. (a) Preparation of coherent resource states by implementing sLOCC on indistinguishable particles. Photon pairs with opposite polarizations are prepared by pumping a $\beta$-barium borate (BBO) crystal. The two photon wavefunctions are distributed in two spatial regions, with the indistinguishability tuned by the half-wave plates (HWPs) \#1 and \#2. The elements in the dashed box are inserted only to prepare the three-level state. (b) Discrimination of different phases. The Franson interferometer creates two phase channels with different configurations, which is adjusted by the HWP sandwiched between two quarter-wave plates (QWPs). Inset: the polarization analysis device (PAD) comprising a QWP, a HWP, a polarizing beam splitter (PBS) and a single-photon detector. BS beam splitter, BD beam displacer.
  }
  \label{fig:setup}
\end{figure*}

\textit{A photonic coherence synthesizer.}---We demonstrate the preparation of two-level and three-level indistinguishability-based coherence by means of the photonic configuration shown in 
Fig.~\ref{fig:setup}(a). The correspondence between photon's polarization and pseudospin reads $\ket{H}\leftrightarrow\ket{\uparrow}$, $\ket{V}\leftrightarrow\ket{\downarrow}$, with $\ket{H}$ and $\ket{V}$ identifying horizontal and vertical polarization, respectively. Frequency-degenerate photon pairs are generated by pumping a beamlike type-II $\beta$-barium borate (BBO) crystal via spontaneous parametric down-conversion \cite{type2}, and sent to the main setup via two single-mode fibers, respectively. The two-photon initial state $\ket{H}\otimes\ket{V}$ is uncorrelated, and two half-wave plates (HWPs, \#1 and \#2) with their orientation set at $22.5^\circ$ and $\theta/2$, respectively, are utilized to adjust their polarization. The two-level state $\ket{\Psi_\mathrm{LR}}$ is effectively prepared by the setup already employed to demonstrate polarization-entanglement activation by spatial indistinguishability \cite{sun2020}. Each of the two initially uncorrelated photons passes through a polarizing beam splitter (PBS), which distributes their spatial wavefunction between two remote sites, L and R, according to the polarization state. Next, additional HWPs are added in different paths to revert the photons' polarization, and a beam displacer (BD) is inserted on each site to combine the propagating directions of the two photons. At this point, the spatial wavefunctions of the two photons become overlapped, allowing for preparation of the state $\ket{\Psi_\mathrm{LR}}$ via sLOCC. Explicitly, a pair of polarization analysis devices (PADs) are inserted to cast polarization measurement, and the coincidence photon counting process realizes the desired projection onto the distributed basis 
$\mathcal{B}$ (for more details, see Ref.~\cite{sun2020}).

To prepare the three-level state $\ket{\Phi_\mathrm{LR}}$, an additional part of setup consisting of an HWP set at $22.5^\circ$ and a BD is appended on each site, L and R, the orientations of the HWPs \#3 and \#4 being also adjusted to prepare one of the photons in the polarization-superposition state (see dashed box in Fig.~\ref{fig:setup}(a)). The coherence underpinning the system is finally activated and detected via sLOCC.

\begin{figure}[t]
  \centering
  \includegraphics[width=0.48\textwidth]{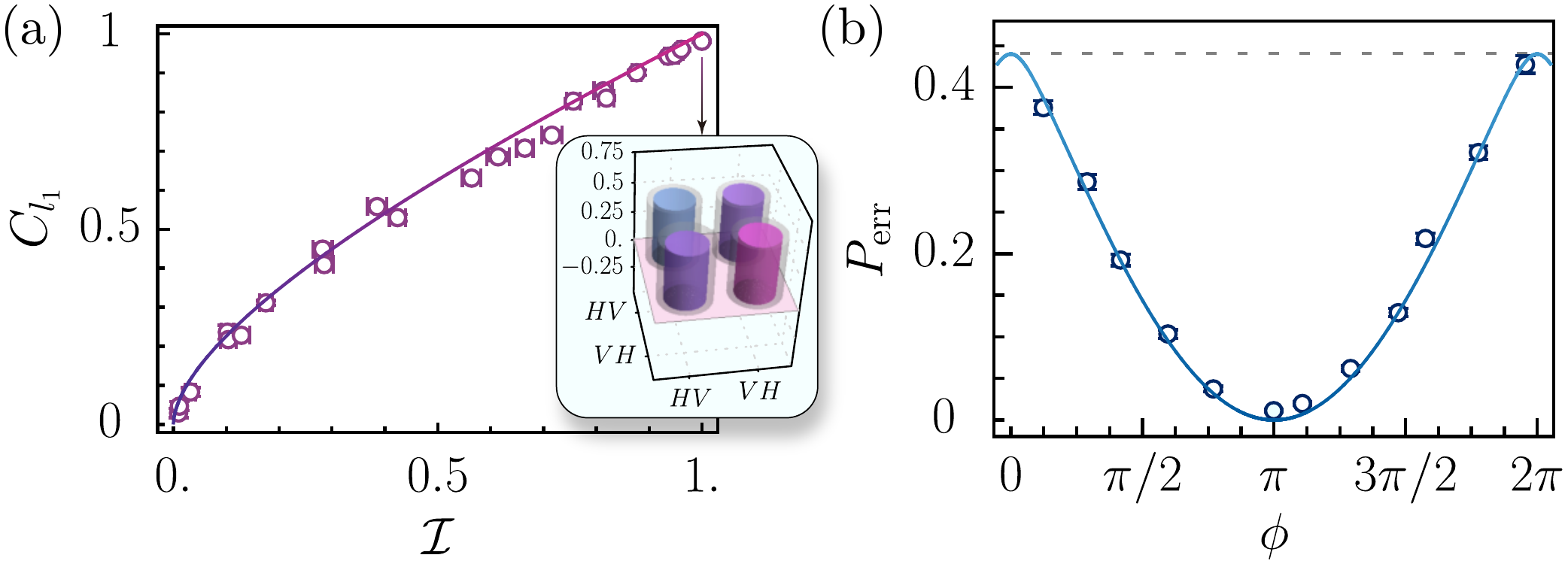}
  \caption{Experimental result for the two-level state $\ket{\Psi_\mathrm{LR}}$. (a) Quantification of coherence $C_{l_1}$ versus the two-photon indistinguishability $\mathcal{I}$. The inset shows the real part of the density matrix for the input state $\ket{\Psi_\mathrm{LR}(\pi/4)}$ deduced by quantum state tomography. (b) The error probability $P_\mathrm{err}$ of phase discrimination versus the phase parameter $\phi$, with $\theta=\pi/4$ to give maximal coherence and $p_1=0.44$. The dashed line shows the Helstrom-Holevo bound without coherence.}
  \label{fig:result2}
\end{figure}

As a first observation, we want to prove the direct quantitative connection between produced coherence and spatial indistinguishability of photons, in analogy to what has been done for the entanglement \cite{sun2020}. In fact, in the present experimental study, the resource of interest is quantum coherence and such a preliminary analysis is essential in view of its controllable exploitation for the specific quantum metrology protocol. This analysis is performed for the two-level state $\ket{\Psi_\mathrm{LR}}$ resulting from the original elementary state $\ket{\Psi}$. Various methods have been proposed to quantify coherence \cite{plenio2014, piani2016, napoli2016, spekkens2016, wang2017}. Here, we adopt the $l_1$ norm of the density matrix $\rho$, that is $C_{l_1}(\rho)=\sum_{i\neq j}|\rho_{ij}|$ \cite{plenio2014}. For convenience, we omit the site delimiters of the distributed state and simply denote it using polarization. The system is prepared in $\ket{\Psi_\mathrm{LR}(\theta)}=\cos\theta\ket{HV}+\sin\theta\ket{VH}$, and its measure of coherence in the basis $\mathcal{B}$ is $C_{l_1}(\Psi_\mathrm{LR})=\left|\,\sin2\theta\,\right|$. The coherence completely stems from the indistinguishability of the photons, as it vanishes at the limit $\theta = k\pi/2$ ($k$ integer number), i.e., when the two photons are distinguishable. To quantify the spatial indistinguishability of the two photons we use the entropic measure~\cite{saro2020npj} $\mathcal{I}=-\sum_{i=1}^2p_\mathrm{LR}^{(i)}\log p_\mathrm{LR}^{(i)}$, where $p_\mathrm{LR}^{(1)}=|lr'/\mathcal{N}_\mathrm{LR}^\Phi|^2$ $(p_\mathrm{LR}^{(2)}=|l'r/\mathcal{N}_\mathrm{LR}^\Phi|^2)$ refers to the probability of finding the photon from $\psi$ and $\psi'$ ($\psi'$ and $\psi$) ending at L and R, respectively. For our setup, one has $\mathcal{I}=-\cos^2\theta \log (\cos^2 \theta)-\sin^2\theta \log (\sin^2 \theta)$. The experimental result for the measurement of coherence versus indistinguishability is plotted in Fig.~\ref{fig:result2}(a), clearly revealing the monotonic dependence in accord with theoretical predictions. Here and after, the error bars represent the $1\sigma$ standard deviation of data points, which is deduced by assuming a Poisson distribution for counting statistics, and resampling over recorded data. The inset shows the result of quantum state tomography at $\theta=\pi/4$, which has a fidelity of 0.988 to the maximally coherent state. 

\textit{Phase discrimination.}---Having generated tunable coherence using sLOCC, we apply it in the phase discrimination task to demonstrate the operational advantage due to indistinguishability and the role of particle statistics. The formal definition of phase discrimination task is as follows: a phase unitary among $n$ possible choices $U_k=e^{i\hat{G}\phi_k},k\in\{1,\ldots,n\}$ is randomly applied on an initial state $\rho_\mathrm{in}$ with a probability of $p_k$, where the generator of the transformation  $\hat{G}=\sum_{\sigma\tau=\uparrow, \downarrow}\omega_{\sigma\tau}\ket{\rm L\sigma,R\tau}\bra{\rm L\uparrow,R\downarrow}$ is diagonal on the computational basis ($\omega_{\sigma\tau}$ are arbitrary coefficients) and $\sum_{k=1}^n p_k=1$. We shall identify the $\phi_k$ that is actually applied with maximal confidence from the output state $\rho_\mathrm{out}$, by casting on it positive operator-valued measurements (POVMs). Here, we focus on the $n=2$ scenario with $\phi_1=0$, $\phi_2=\phi$, and solving the task using the experimentally feasible minimum-error discrimination \cite{croke2009, mosley2006}.

We first investigate phase discrimination with the two-level state and, without loss of generality, choose the generator $\hat{G}=\ket{\rm L\uparrow,R\downarrow}\bra{\rm L\uparrow,R\downarrow}$ (obtained fixing $\omega_{\uparrow\downarrow}=1$ and $\omega_{\uparrow\uparrow}=\omega_{\downarrow\uparrow}=\omega_{\downarrow\downarrow}=0$). Consequently, the output states after being affected by $U_k$ read
\begin{align}\label{eq:psik}
    \ket{\Psi^k}=\frac{1}{\mathcal{N}_\mathrm{LR}^\Psi}(lr'\ket{\rm L\downarrow,R\uparrow}+\eta l're^{i(k-1)\phi}\ket{\rm L\uparrow,R\downarrow}),
\end{align}
and they are discriminated by a POVM (a von Neumann projective measurement in this case) comprising two projectors $\mathbf{\Pi}=\{\hat{\Pi}_1,\hat{\Pi}_2\}$: when $\hat{\Pi}_k$ clicks, the phase is identified as $\phi_k$. By this definition, the chance of making an error is $P_\mathrm{err}=p_1\langle\Psi^1|\hat{\Pi}_2|\Psi^1\rangle+p_2\langle\Psi^2|\hat{\Pi}_1|\Psi^2\rangle$, and is lower bounded by the Helstrom-Holevo bound \cite{Helstrom,Holevo}, namely, $P_\mathrm{err}\geqslant\frac{1}{2}\left(1-\sqrt{1-4p_1p_2\left|\braket{\Psi^1|\Psi^2}\right|^2}\right)$. For a two-level coherent state, it is straightforward to identify the measurement projectors $\hat{\Pi}_1$ and $\hat{\Pi}_2$ \cite{saro2019pra}.

The phase discrimination game is experimentally realized using the setup of Fig.~\ref{fig:setup}(b). The photons in the state $\ket{\Psi_\mathrm{LR}}$ on the site R are sent into a unbalanced Mach-Zehnder interferometer (UMZI), while the photons on the site L are directly detected. We put a HWP between two QWPs fixed at $45^\circ$ to build a phase gate, and situate one phase gate into each of the arms after a non-polarization beam splitter (BS). In fact, in the short arm of UMZI, the state $\ket{\Psi_\mathrm{LR}}$ remains unchanged, while in the long arm, a relative phase $\phi$ between $\ket{{\rm L} V,\ {\rm R} H}$ and $\ket{{\rm L} H,\ {\rm R} V}$ is imported. A movable shutter (not shown) is placed in one of the arms to adjust the parameters $p_1$ and $p_2$. After the MZI, the photons are projected on the desired state. Since $\ket{\Psi_\mathrm{LR}}$ is a two-level coherent state, the measurement projectors $\hat{\Pi}_1$ and $\hat{\Pi}_2$ defined in the basis $\{\ket{{\rm L} V,\ {\rm R} H},\ \ket{{\rm L} H,\ {\rm R} V}\}$ are realized by drawing the corresponding subspace from the product (single-particle) state measurement. This procedure is as follows. On the site L (R), the polarization projector is $\hat{O}_\mathrm{L}=\ket{\chi}\bra{\chi}$ with $\ket{\chi}=\alpha\ket{H}+\beta\ket{V}$ ($\hat{O}'_\mathrm{R}=\ket{\chi'}\bra{\chi'}$ with $\ket{\chi'}=\alpha'\ket{H}+\beta'\ket{V}$); the product projector is thus $\hat{O}_\mathrm{L}\otimes \hat{O}'_\mathrm{R}$, leading to the two-photon projector $\ket{\Psi_{\alpha \beta}}\bra{\Psi_{\alpha \beta}}$ with $\ket{\Psi_{\alpha\beta}}=\alpha\beta'\ket{{\rm L} H,\ {\rm R} V} + \beta\alpha'\ket{{\rm L} V,\ {\rm R} H}$ in the subspace of interest $\{\ket{{\rm L} V,\ {\rm R} H},\ \ket{{\rm L} H,\ {\rm R} V}\}$. Thanks to the final PAD unit of the setup of Fig.~\ref{fig:setup}(b), the parameters $\{\alpha,\ \beta,\ \alpha',\ \beta'\}$ can be adjusted to perform the desired projective measurements $\hat{\Pi}_1$, $\hat{\Pi}_2$ and eventually obtain the error probability of discrimination $P_\mathrm{err}$. 

We directly measure the error probability of phase discrimination for various $\phi$ at $p_1=0.44$ by employing the maximally coherent state $\ket{\Psi_\mathrm{LR}(\pi/4)}$ and optimizing over the measurement settings of $\hat{\Pi}_1$ and $\hat{\Pi}_2$. The experimental result, matching well with the theoretical prediction,
\begin{align}\label{eq:perr}
    P_\mathrm{err}=\frac{1}{2}\left(1-\sqrt{1-2p_1(1-p_1)(1+\cos\phi)}\right),
\end{align}
is shown in Fig.~\ref{fig:result2}(b). Note that without coherence, the best strategy of phase discrimination is to constantly guess the phase with greater probability, yielding $\bar{P}_\mathrm{err}=p_1$ (top dashed line). The reduced $P_\mathrm{err}$ thus unravels the almost ubiquitous advantage of indistinguishability-based coherence.

\textit{Particle statistics matters.}---The symmetric form of Eq.~(\ref{eq:psik}) prevents the exchange phase factor $\eta$ from affecting the outcome of $\ket{\Psi_\mathrm{LR}}$-based phase discrimination task. However, when $\ket{\Phi_\mathrm{LR}}$ is utilized in the same task, the intrinsic statistics of the indistinguishable particles renders the situation more complicated. In our optical setup, any state prepared necessarily has $\eta=+1$. For simplicity, we choose $a=b$ and set $l=l'=r=r'$ $(l'=r=0)$ to maximize (destroy) indistinguishability. This is experimentally achieved by setting the orientation of HWPs \#3 and \#4 be $22.5^\circ$ and $\theta=\pi/4(0)$. However, investigation of Fermionic systems with $\eta=-1$ is also possible, which follows from the observation that $\eta$ in Eq.~\ref{eq:phi} can be absorbed into $l'$. By setting $\theta=-\pi/4$, we invert the sign of $l'$ to simulate indistinguishability-activated coherence of Fermionic particles.

\begin{figure}[t!]
  \centering
  \includegraphics[width=0.48\textwidth]{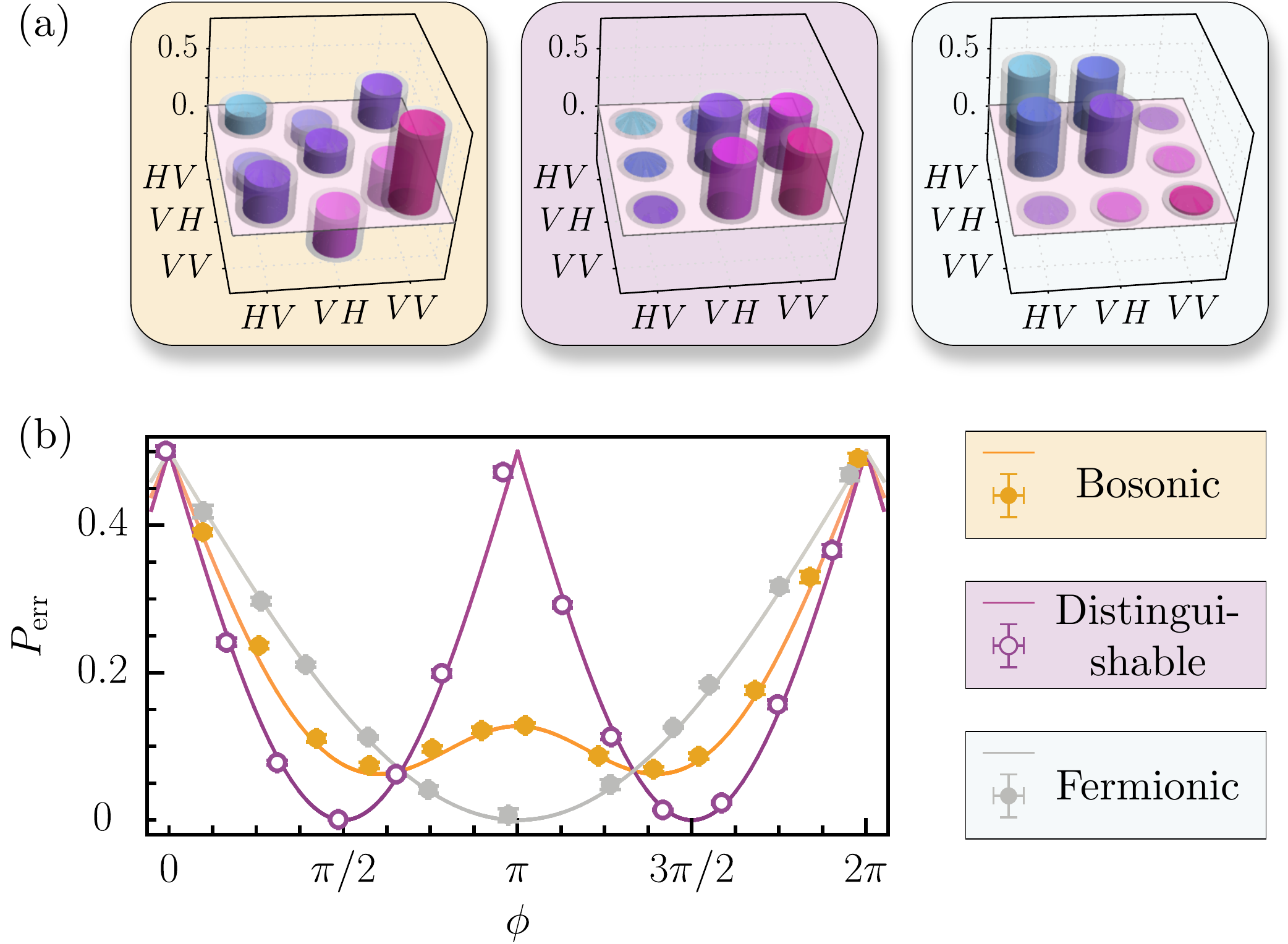}
  \caption{Experimental result for three-level state $\ket{\Phi_\mathrm{LR}}$ with $p_1=0.50$. (a) The real part of the density matrix for the input states $\ket{\Phi_\mathrm{LR}}$ of Bosonic, distinguishable and Fermionic particles (simulated), deduced by quantum state tomography, with $\theta=\pm\pi/4$ to give maximal coherence. (b) The error probability $P_\mathrm{err}$ of phase discrimination versus $\phi$. 
  }
  \label{fig:result3}
\end{figure}

The state preparation in all above cases is characterized via quantum state tomography, and the results are presented in Fig.~\ref{fig:result3}(a), the magnitude of the imaginary part of the density matrices are smaller than 0.07. For the Bosonic case, the outcome authenticates the presence of coherence between all three vectors of the computational basis shown in Eq.~(\ref{eq:phi}). For the distinguishable case, the coherence is in contrast solely inherited from one of the particles, and localized on the site R. For the Fermionic case, the destructive interference completely eliminates the amplitude on the symmetric basis $\ket{VV}\sim\ket{\rm L\downarrow,R\downarrow}$, and the resulted state interestingly resembles the two-level state, $\ket{\Psi_\mathrm{LR}(\pi/4)}$, as is substantiated by the deduced density matrices. In the experiment, we simulate this case with the average fidelity of $95.9(2)\%$ compared with $\ket{\Psi_\mathrm{LR}(\pm\pi/4)}$. Notice that a minus sign appears in the coefficient of the $\ket{VH}$ terms, which is attributed to the $\pi$-phase acquired by the photons upon reflected by PBS.

We are now in the position to demonstrate the role of particle statistics in the phase discrimination task. The corresponding operations $U_k$ are again realized using the phase gates within the UMZI, yielding two output states $\ket{\Phi^k}$ \cite{saro2019pra} written as
\begin{align}\label{eq:phik}
   \ket{\Phi^k}=(&a(lr'e^{i\omega_{\downarrow\uparrow}\phi_k} \ket{L\downarrow, R\uparrow}+\eta l' r e^{i\omega_{\uparrow\downarrow}\phi_k} \ket{L\uparrow, R\downarrow}) \nonumber\\
  +&b (l r'+ \eta l' r) e^{i\omega_{\downarrow\downarrow}\phi_k} \ket{L\downarrow, R\downarrow})/\mathcal{N}_{LR}^\Phi.
 \end{align}
Here, we set  $\omega_{\downarrow\uparrow}=1$, $\omega_{\uparrow\downarrow}=2$ and $\omega_{\downarrow\downarrow}=3$ in the generator $\hat{G}$.
Differently from the two-level situation, in this three-level coherent case, we need to place each UMZI on each site L and R. The UMZI has a path difference equivalent to 2.7ns between the long and short paths, and the coincidence interval is set at 0.8ns. The quantum states affected by the two phase operations in the UMZIs are registered separately \cite{barnett1997, mohseni2004}. We adjust the electronic delay of the coincidence module to pick out the events that the two photons had taken the long/short and short/long paths, which correspond to the state after being affected by $U_1\text{~and~}U_2$, respectively.
Moreover, for the measurement of the three-level system, to minimize the error probability of discrimination $P_\mathrm{err}$, three projectors $\hat{\Pi}_1$, $\hat{\Pi}_2$ and $\hat{\Pi}_3$ are required where $\Sigma_i^3\hat{\Pi}_i=I$ and $\rm{Tr}[\hat{\Pi}_3\ket{\Phi_{LR}^1}\bra{\Phi_{LR}^1}]=\rm{Tr}[\hat{\Pi}_3\ket{\Phi_{LR}^2}\bra{\Phi_{LR}^2}]=0$. The projectors $\hat{\Pi}_i$ ($i=1,\ 2,\ 3$) consist of three linearly independent basis vectors $\mathcal{B}'=\{\ket{\rm L\uparrow,R\downarrow},\ket{\rm L\downarrow,R\uparrow},\ket{\rm L\downarrow,R\downarrow}\}$. Similarly to the method used above for the two-level state, these three projectors are also extracted from the subspace of the product projectors on the two sites L and R and implemented by the PAD unit of the setup.


Fig.~\ref{fig:result3}(b) reports the measured error probabilities for phase discrimination with the three-level states, where we omit the experimental result for the Fermionic case, because it is identical to the two-level case already given in the earlier text (see Fig.~\ref{fig:result2}(b)). A clear discrepancy between the credibility of phase discrimination using different kinds of particles can be observed. Particularly, both types of indistinguishable particles provide advantage over distinguishable ones within the range of $\phi\in(\frac{2\pi}{3},\frac{4\pi}{3})$, but  Fermions further outperform Bosons by a difference in $P_\mathrm{err}$ of 0.119 at $\phi=\pi$. This can be intuitively interpreted by recalling that the exchange interaction of Fermions prevent them from occupying the same state, so the wavefunction amplitude disperses between different states and produces large amount of coherence. In contrast, Bosons tend to bunch on a single state, so the applicable coherence is reduced.

\textit{Discussion.}---Coherence activated from spatial indistinguishability is a fundamental contribution to quantumness of multiparticle composite systems intimately related to the presence of identical particles (subsystems). It cannot exist between different types of quanta, that is, in systems made of nonidentical (or distinguishable) particles. Due to its intrinsic nonlocal trait, in order to exploit indistinguishability-based coherence for quantum information tasks, transformations and measurements on the resource state must admit direct product decomposition into local operations, which are achieved by sLOCC. We note that in the case of two identical particles, Schmidt decomposition recovers our capability to perform all possible measurements \cite{SciaraSciRep}. Therefore, the application of indistinguishability-based coherence between three or more quanta will be an open research route.

In this paper, we have experimentally investigated indistinguishability-based coherence, demonstrating its operational usefulness in a quantum metrology protocol. Our photonic architecture is capable of tuning the degree of spatial indistinguishability of two uncorrelated photons, and adjusting the interplay between indistinguishability-based coherence and single-particle superposition-based coherence. This has allowed us to prepare via sLOCC various types of resource states and exploit them in the phase discrimination task to characterize the operational coherence. Interestingly, our setup has been designed in such a way that both Bosonic and Fermionic statistics can occur in the resource states, thus enabling the possibility to directly observe how the nature of the employed particles affects the efficiency of the quantum task. 
Our results highlight, in a comprehensive fashion, the fundamental and practical aspects of controllable indistinguishability of identical building blocks for quantum-enhanced technologies.

\begin{acknowledgments}
%

This work was supported by
National Key Research and Development Program of China (Grants
Nos.\,2016YFA0302700, 2017YFA0304100),
the National Natural Science Foundation of China (Grants
Nos.\,61725504, U19A2075,
61805227, 61975195,
11774335, and 11821404),
Key Research Program of Frontier Sciences, CAS (Grant No.\,QYZDY-SSW-SLH003),
Science Foundation of the CAS (Grant No.\,ZDRW-XH-2019-1),
the Fundamental Research Funds for the Central Universities (Grant No.\,WK2470000026, No.\,WK2030380017),  
Anhui Initiative in Quantum Information Technologies (Grants No.\,AHY020100, and No.\,AHY060300). L.L. acknowledges support from the Alexander von Humboldt Foundation.

K.S. and Z.-H.L. contributed equally to this work. 

\end{acknowledgments}


\begin{thebibliography}{41}%
\makeatletter
\providecommand \@ifxundefined [1]{%
 \@ifx{#1\undefined}
}%
\providecommand \@ifnum [1]{%
 \ifnum #1\expandafter \@firstoftwo
 \else \expandafter \@secondoftwo
 \fi
}%
\providecommand \@ifx [1]{%
 \ifx #1\expandafter \@firstoftwo
 \else \expandafter \@secondoftwo
 \fi
}%
\providecommand \natexlab [1]{#1}%
\providecommand \enquote  [1]{``#1''}%
\providecommand \bibnamefont  [1]{#1}%
\providecommand \bibfnamefont [1]{#1}%
\providecommand \citenamefont [1]{#1}%
\providecommand \href@noop [0]{\@secondoftwo}%
\providecommand \href [0]{\begingroup \@sanitize@url \@href}%
\providecommand \@href[1]{\@@startlink{#1}\@@href}%
\providecommand \@@href[1]{\endgroup#1\@@endlink}%
\providecommand \@sanitize@url [0]{\catcode `\\12\catcode `\$12\catcode
  `\&12\catcode `\#12\catcode `\^12\catcode `\_12\catcode `\%12\relax}%
\providecommand \@@startlink[1]{}%
\providecommand \@@endlink[0]{}%
\providecommand \url  [0]{\begingroup\@sanitize@url \@url }%
\providecommand \@url [1]{\endgroup\@href {#1}{\urlprefix }}%
\providecommand \urlprefix  [0]{URL }%
\providecommand \Eprint [0]{\href }%
\providecommand \doibase [0]{https://doi.org/}%
\providecommand \selectlanguage [0]{\@gobble}%
\providecommand \bibinfo  [0]{\@secondoftwo}%
\providecommand \bibfield  [0]{\@secondoftwo}%
\providecommand \translation [1]{[#1]}%
\providecommand \BibitemOpen [0]{}%
\providecommand \bibitemStop [0]{}%
\providecommand \bibitemNoStop [0]{.\EOS\space}%
\providecommand \EOS [0]{\spacefactor3000\relax}%
\providecommand \BibitemShut  [1]{\csname bibitem#1\endcsname}%
\let\auto@bib@innerbib\@empty
\bibitem [{\citenamefont {Gleason}(1957)}]{gleason1957}%
  \BibitemOpen
  \bibfield  {author} {\bibinfo {author} {\bibfnamefont {A.~M.}\ \bibnamefont
  {Gleason}},\ }\href {https://www.jstor.org/stable/24900629} {\bibfield
  {journal} {\bibinfo  {journal} {J. Math. Mech.}\ }\textbf {\bibinfo {volume}
  {6}},\ \bibinfo {pages} {885} (\bibinfo {year} {1957})}\BibitemShut {NoStop}%
\bibitem [{\citenamefont {Kochen}\ and\ \citenamefont
  {Specker}(1967)}]{ks1967}%
  \BibitemOpen
  \bibfield  {author} {\bibinfo {author} {\bibfnamefont {S.}~\bibnamefont
  {Kochen}}\ and\ \bibinfo {author} {\bibfnamefont {E.}~\bibnamefont
  {Specker}},\ }\href {https://www.jstor.org/stable/24902153} {\bibfield
  {journal} {\bibinfo  {journal} {J. Math. Mech.}\ }\textbf {\bibinfo {volume}
  {17}},\ \bibinfo {pages} {59} (\bibinfo {year} {1967})}\BibitemShut {NoStop}%
\bibitem [{\citenamefont {Schr{\"o}dinger}(1935)}]{schrodinger}%
  \BibitemOpen
  \bibfield  {author} {\bibinfo {author} {\bibfnamefont {E.}~\bibnamefont
  {Schr{\"o}dinger}},\ }\href@noop {} {\bibfield  {journal} {\bibinfo
  {journal} {Naturwissenschaften}\ }\textbf {\bibinfo {volume} {23}},\ \bibinfo
  {pages} {823} (\bibinfo {year} {1935})}\BibitemShut {NoStop}%
\bibitem [{\citenamefont {Wigner}(1995)}]{wigner}%
  \BibitemOpen
  \bibfield  {author} {\bibinfo {author} {\bibfnamefont {E.~P.}\ \bibnamefont
  {Wigner}},\ }in\ \href@noop {} {\emph {\bibinfo {booktitle} {Philosophical
  reflections and syntheses}}}\ (\bibinfo  {publisher} {Springer},\ \bibinfo
  {year} {1995})\ pp.\ \bibinfo {pages} {247--260}\BibitemShut {NoStop}%
\bibitem [{\citenamefont {Frauchiger}\ and\ \citenamefont
  {Renner}(2018)}]{renner2018}%
  \BibitemOpen
  \bibfield  {author} {\bibinfo {author} {\bibfnamefont {D.}~\bibnamefont
  {Frauchiger}}\ and\ \bibinfo {author} {\bibfnamefont {R.}~\bibnamefont
  {Renner}},\ }\href {https://doi.org/10.1038/s41467-018-05739-8} {\bibfield
  {journal} {\bibinfo  {journal} {Nat. Commun.}\ }\textbf {\bibinfo {volume}
  {9}},\ \bibinfo {pages} {3711} (\bibinfo {year} {2018})}\BibitemShut
  {NoStop}%
\bibitem [{\citenamefont {Bong}\ \emph {et~al.}(2020)\citenamefont {Bong},
  \citenamefont {Utreras-Alarc{\'o}n}, \citenamefont {Ghafari}, \citenamefont
  {Liang}, \citenamefont {Tischler}, \citenamefont {Cavalcanti}, \citenamefont
  {Pryde},\ and\ \citenamefont {Wiseman}}]{wiseman2020}%
  \BibitemOpen
  \bibfield  {author} {\bibinfo {author} {\bibfnamefont {K.-W.}\ \bibnamefont
  {Bong}}, \bibinfo {author} {\bibfnamefont {A.}~\bibnamefont
  {Utreras-Alarc{\'o}n}}, \bibinfo {author} {\bibfnamefont {F.}~\bibnamefont
  {Ghafari}}, \bibinfo {author} {\bibfnamefont {Y.-C.}\ \bibnamefont {Liang}},
  \bibinfo {author} {\bibfnamefont {N.}~\bibnamefont {Tischler}}, \bibinfo
  {author} {\bibfnamefont {E.~G.}\ \bibnamefont {Cavalcanti}}, \bibinfo
  {author} {\bibfnamefont {G.~J.}\ \bibnamefont {Pryde}},\ and\ \bibinfo
  {author} {\bibfnamefont {H.~M.}\ \bibnamefont {Wiseman}},\ }\href
  {https://doi.org/10.1038/s41567-020-0990-x} {\bibfield  {journal} {\bibinfo
  {journal} {Nat. Phys.}\ }\textbf {\bibinfo {volume} {16}},\ \bibinfo {pages}
  {1199} (\bibinfo {year} {2020})}\BibitemShut {NoStop}%
\bibitem [{\citenamefont {Bera}\ \emph {et~al.}(2017)\citenamefont {Bera},
  \citenamefont {Ac{\'{\i}}n}, \citenamefont {Ku{\'{s}}}, \citenamefont
  {Mitchell},\ and\ \citenamefont {Lewenstein}}]{Bera_2017}%
  \BibitemOpen
  \bibfield  {author} {\bibinfo {author} {\bibfnamefont {M.~N.}\ \bibnamefont
  {Bera}}, \bibinfo {author} {\bibfnamefont {A.}~\bibnamefont {Ac{\'{\i}}n}},
  \bibinfo {author} {\bibfnamefont {M.}~\bibnamefont {Ku{\'{s}}}}, \bibinfo
  {author} {\bibfnamefont {M.~W.}\ \bibnamefont {Mitchell}},\ and\ \bibinfo
  {author} {\bibfnamefont {M.}~\bibnamefont {Lewenstein}},\ }\href
  {https://doi.org/10.1088/1361-6633/aa8731} {\bibfield  {journal} {\bibinfo
  {journal} {Rep. Prog. Phys.}\ }\textbf {\bibinfo {volume} {80}},\ \bibinfo
  {pages} {124001} (\bibinfo {year} {2017})}\BibitemShut {NoStop}%
\bibitem [{\citenamefont {Shor}(1994)}]{shor1994}%
  \BibitemOpen
  \bibfield  {author} {\bibinfo {author} {\bibfnamefont {P.~W.}\ \bibnamefont
  {Shor}},\ }in\ \href@noop {} {\emph {\bibinfo {booktitle} {Proceedings 35th
  annual symposium on foundations of computer science}}}\ (\bibinfo
  {organization} {Ieee},\ \bibinfo {year} {1994})\ pp.\ \bibinfo {pages}
  {124--134}\BibitemShut {NoStop}%
\bibitem [{\citenamefont {Martin-Lopez}\ \emph {et~al.}(2012)\citenamefont
  {Martin-Lopez}, \citenamefont {Laing}, \citenamefont {Lawson}, \citenamefont
  {Alvarez}, \citenamefont {Zhou},\ and\ \citenamefont {O'brien}}]{obrien2012}%
  \BibitemOpen
  \bibfield  {author} {\bibinfo {author} {\bibfnamefont {E.}~\bibnamefont
  {Martin-Lopez}}, \bibinfo {author} {\bibfnamefont {A.}~\bibnamefont {Laing}},
  \bibinfo {author} {\bibfnamefont {T.}~\bibnamefont {Lawson}}, \bibinfo
  {author} {\bibfnamefont {R.}~\bibnamefont {Alvarez}}, \bibinfo {author}
  {\bibfnamefont {X.-Q.}\ \bibnamefont {Zhou}},\ and\ \bibinfo {author}
  {\bibfnamefont {J.~L.}\ \bibnamefont {O'brien}},\ }\href
  {https://doi.org/10.1038/nphoton.2012.259} {\bibfield  {journal} {\bibinfo
  {journal} {Nat. Photon.}\ }\textbf {\bibinfo {volume} {6}},\ \bibinfo {pages}
  {773} (\bibinfo {year} {2012})}\BibitemShut {NoStop}%
\bibitem [{\citenamefont {Cleve}\ \emph {et~al.}(1998)\citenamefont {Cleve},
  \citenamefont {Ekert}, \citenamefont {Macchiavello},\ and\ \citenamefont
  {Mosca}}]{cleve1998}%
  \BibitemOpen
  \bibfield  {author} {\bibinfo {author} {\bibfnamefont {R.}~\bibnamefont
  {Cleve}}, \bibinfo {author} {\bibfnamefont {A.}~\bibnamefont {Ekert}},
  \bibinfo {author} {\bibfnamefont {C.}~\bibnamefont {Macchiavello}},\ and\
  \bibinfo {author} {\bibfnamefont {M.}~\bibnamefont {Mosca}},\ }\href
  {https://doi.org/10.1098/rspa.1998.0164} {\bibfield  {journal} {\bibinfo
  {journal} {Proc. R. Soc. Lond. A}\ }\textbf {\bibinfo {volume} {454}},\
  \bibinfo {pages} {339} (\bibinfo {year} {1998})}\BibitemShut {NoStop}%
\bibitem [{\citenamefont {Procopio}\ \emph {et~al.}(2015)\citenamefont
  {Procopio}, \citenamefont {Moqanaki}, \citenamefont {Ara{\'u}jo},
  \citenamefont {Costa}, \citenamefont {Calafell}, \citenamefont {Dowd},
  \citenamefont {Hamel}, \citenamefont {Rozema}, \citenamefont {Brukner},\ and\
  \citenamefont {Walther}}]{procopio2015}%
  \BibitemOpen
  \bibfield  {author} {\bibinfo {author} {\bibfnamefont {L.~M.}\ \bibnamefont
  {Procopio}}, \bibinfo {author} {\bibfnamefont {A.}~\bibnamefont {Moqanaki}},
  \bibinfo {author} {\bibfnamefont {M.}~\bibnamefont {Ara{\'u}jo}}, \bibinfo
  {author} {\bibfnamefont {F.}~\bibnamefont {Costa}}, \bibinfo {author}
  {\bibfnamefont {I.~A.}\ \bibnamefont {Calafell}}, \bibinfo {author}
  {\bibfnamefont {E.~G.}\ \bibnamefont {Dowd}}, \bibinfo {author}
  {\bibfnamefont {D.~R.}\ \bibnamefont {Hamel}}, \bibinfo {author}
  {\bibfnamefont {L.~A.}\ \bibnamefont {Rozema}}, \bibinfo {author}
  {\bibfnamefont {{\v{C}}.}~\bibnamefont {Brukner}},\ and\ \bibinfo {author}
  {\bibfnamefont {P.}~\bibnamefont {Walther}},\ }\href
  {https://doi.org/10.1038/ncomms8913} {\bibfield  {journal} {\bibinfo
  {journal} {Nat. Commun.}\ }\textbf {\bibinfo {volume} {6}},\ \bibinfo {pages}
  {7913} (\bibinfo {year} {2015})}\BibitemShut {NoStop}%
\bibitem [{\citenamefont {Aaronson}\ and\ \citenamefont
  {Arkhipov}(2011)}]{aaronson2011}%
  \BibitemOpen
  \bibfield  {author} {\bibinfo {author} {\bibfnamefont {S.}~\bibnamefont
  {Aaronson}}\ and\ \bibinfo {author} {\bibfnamefont {A.}~\bibnamefont
  {Arkhipov}},\ }in\ \href {https://doi.org/10.1145/1993636.1993682} {\emph
  {\bibinfo {booktitle} {Proceedings of the forty-third annual ACM symposium on
  Theory of computing}}}\ (\bibinfo {year} {2011})\ pp.\ \bibinfo {pages}
  {333--342}\BibitemShut {NoStop}%
\bibitem [{\citenamefont {Arute}\ \emph {et~al.}(2019)\citenamefont {Arute},
  \citenamefont {Arya}, \citenamefont {Babbush}, \citenamefont {Bacon},
  \citenamefont {Bardin}, \citenamefont {Barends}, \citenamefont {Biswas},
  \citenamefont {Boixo}, \citenamefont {Brandao}, \citenamefont {Buell} \emph
  {et~al.}}]{google2019}%
  \BibitemOpen
  \bibfield  {author} {\bibinfo {author} {\bibfnamefont {F.}~\bibnamefont
  {Arute}}, \bibinfo {author} {\bibfnamefont {K.}~\bibnamefont {Arya}},
  \bibinfo {author} {\bibfnamefont {R.}~\bibnamefont {Babbush}}, \bibinfo
  {author} {\bibfnamefont {D.}~\bibnamefont {Bacon}}, \bibinfo {author}
  {\bibfnamefont {J.~C.}\ \bibnamefont {Bardin}}, \bibinfo {author}
  {\bibfnamefont {R.}~\bibnamefont {Barends}}, \bibinfo {author} {\bibfnamefont
  {R.}~\bibnamefont {Biswas}}, \bibinfo {author} {\bibfnamefont
  {S.}~\bibnamefont {Boixo}}, \bibinfo {author} {\bibfnamefont {F.~G.}\
  \bibnamefont {Brandao}}, \bibinfo {author} {\bibfnamefont {D.~A.}\
  \bibnamefont {Buell}}, \emph {et~al.},\ }\href
  {https://doi.org/10.1038/s41586-019-1666-5} {\bibfield  {journal} {\bibinfo
  {journal} {Nature}\ }\textbf {\bibinfo {volume} {574}},\ \bibinfo {pages}
  {505} (\bibinfo {year} {2019})}\BibitemShut {NoStop}%
\bibitem [{\citenamefont {Winter}\ and\ \citenamefont
  {Yang}(2016)}]{andreas2016}%
  \BibitemOpen
  \bibfield  {author} {\bibinfo {author} {\bibfnamefont {A.}~\bibnamefont
  {Winter}}\ and\ \bibinfo {author} {\bibfnamefont {D.}~\bibnamefont {Yang}},\
  }\href {https://doi.org/10.1103/PhysRevLett.116.120404} {\bibfield  {journal}
  {\bibinfo  {journal} {Phys. Rev. Lett.}\ }\textbf {\bibinfo {volume} {116}},\
  \bibinfo {pages} {120404} (\bibinfo {year} {2016})}\BibitemShut {NoStop}%
\bibitem [{\citenamefont {Streltsov}\ \emph {et~al.}(2017)\citenamefont
  {Streltsov}, \citenamefont {Adesso},\ and\ \citenamefont
  {Plenio}}]{adesso2017}%
  \BibitemOpen
  \bibfield  {author} {\bibinfo {author} {\bibfnamefont {A.}~\bibnamefont
  {Streltsov}}, \bibinfo {author} {\bibfnamefont {G.}~\bibnamefont {Adesso}},\
  and\ \bibinfo {author} {\bibfnamefont {M.~B.}\ \bibnamefont {Plenio}},\
  }\href {https://doi.org/10.1103/RevModPhys.89.041003} {\bibfield  {journal}
  {\bibinfo  {journal} {Rev. Mod. Phys.}\ }\textbf {\bibinfo {volume} {89}},\
  \bibinfo {pages} {041003} (\bibinfo {year} {2017})}\BibitemShut {NoStop}%
\bibitem [{\citenamefont {Chitambar}\ and\ \citenamefont
  {Gour}(2019)}]{chitambar2019}%
  \BibitemOpen
  \bibfield  {author} {\bibinfo {author} {\bibfnamefont {E.}~\bibnamefont
  {Chitambar}}\ and\ \bibinfo {author} {\bibfnamefont {G.}~\bibnamefont
  {Gour}},\ }\href {https://doi.org/10.1103/RevModPhys.91.025001} {\bibfield
  {journal} {\bibinfo  {journal} {Rev. Mod. Phys.}\ }\textbf {\bibinfo {volume}
  {91}},\ \bibinfo {pages} {025001} (\bibinfo {year} {2019})}\BibitemShut
  {NoStop}%
\bibitem [{\citenamefont {Castellini}\ \emph
  {et~al.}(2019{\natexlab{a}})\citenamefont {Castellini}, \citenamefont
  {Lo~Franco}, \citenamefont {Lami}, \citenamefont {Winter}, \citenamefont
  {Adesso},\ and\ \citenamefont {Compagno}}]{saro2019pra}%
  \BibitemOpen
  \bibfield  {author} {\bibinfo {author} {\bibfnamefont {A.}~\bibnamefont
  {Castellini}}, \bibinfo {author} {\bibfnamefont {R.}~\bibnamefont
  {Lo~Franco}}, \bibinfo {author} {\bibfnamefont {L.}~\bibnamefont {Lami}},
  \bibinfo {author} {\bibfnamefont {A.}~\bibnamefont {Winter}}, \bibinfo
  {author} {\bibfnamefont {G.}~\bibnamefont {Adesso}},\ and\ \bibinfo {author}
  {\bibfnamefont {G.}~\bibnamefont {Compagno}},\ }\href
  {https://doi.org/10.1103/PhysRevA.100.012308} {\bibfield  {journal} {\bibinfo
   {journal} {Phys. Rev. A}\ }\textbf {\bibinfo {volume} {100}},\ \bibinfo
  {pages} {012308} (\bibinfo {year} {2019}{\natexlab{a}})}\BibitemShut
  {NoStop}%
\bibitem [{\citenamefont {Chin}\ and\ \citenamefont {Huh}(2019)}]{Chin2019}%
  \BibitemOpen
  \bibfield  {author} {\bibinfo {author} {\bibfnamefont {S.}~\bibnamefont
  {Chin}}\ and\ \bibinfo {author} {\bibfnamefont {J.}~\bibnamefont {Huh}},\
  }\href {https://doi.org/10.1103/PhysRevA.99.052345} {\bibfield  {journal}
  {\bibinfo  {journal} {Phys. Rev. A}\ }\textbf {\bibinfo {volume} {99}},\
  \bibinfo {pages} {052345} (\bibinfo {year} {2019})}\BibitemShut {NoStop}%
\bibitem [{\citenamefont {Sperling}\ \emph {et~al.}(2017)\citenamefont
  {Sperling}, \citenamefont {Perez-Leija}, \citenamefont {Busch},\ and\
  \citenamefont {Walmsley}}]{walmsleyPRA2017}%
  \BibitemOpen
  \bibfield  {author} {\bibinfo {author} {\bibfnamefont {J.}~\bibnamefont
  {Sperling}}, \bibinfo {author} {\bibfnamefont {A.}~\bibnamefont
  {Perez-Leija}}, \bibinfo {author} {\bibfnamefont {K.}~\bibnamefont {Busch}},\
  and\ \bibinfo {author} {\bibfnamefont {I.~A.}\ \bibnamefont {Walmsley}},\
  }\href {https://doi.org/10.1103/PhysRevA.96.032334} {\bibfield  {journal}
  {\bibinfo  {journal} {Phys. Rev. A}\ }\textbf {\bibinfo {volume} {96}},\
  \bibinfo {pages} {032334} (\bibinfo {year} {2017})}\BibitemShut {NoStop}%
\bibitem [{\citenamefont {Lo~Franco}\ and\ \citenamefont
  {Compagno}(2018)}]{saro2018prl}%
  \BibitemOpen
  \bibfield  {author} {\bibinfo {author} {\bibfnamefont {R.}~\bibnamefont
  {Lo~Franco}}\ and\ \bibinfo {author} {\bibfnamefont {G.}~\bibnamefont
  {Compagno}},\ }\href {https://doi.org/10.1103/PhysRevLett.120.240403}
  {\bibfield  {journal} {\bibinfo  {journal} {Phys. Rev. Lett.}\ }\textbf
  {\bibinfo {volume} {120}},\ \bibinfo {pages} {240403} (\bibinfo {year}
  {2018})}\BibitemShut {NoStop}%
\bibitem [{\citenamefont {Castellini}\ \emph
  {et~al.}(2019{\natexlab{b}})\citenamefont {Castellini}, \citenamefont
  {Bellomo}, \citenamefont {Compagno},\ and\ \citenamefont
  {Franco}}]{castellini2019pra}%
  \BibitemOpen
  \bibfield  {author} {\bibinfo {author} {\bibfnamefont {A.}~\bibnamefont
  {Castellini}}, \bibinfo {author} {\bibfnamefont {B.}~\bibnamefont {Bellomo}},
  \bibinfo {author} {\bibfnamefont {G.}~\bibnamefont {Compagno}},\ and\
  \bibinfo {author} {\bibfnamefont {R.~L.}\ \bibnamefont {Franco}},\ }\href
  {https://doi.org/10.1103/PhysRevA.99.062322} {\bibfield  {journal} {\bibinfo
  {journal} {Physical Review A}\ }\textbf {\bibinfo {volume} {99}},\ \bibinfo
  {pages} {062322} (\bibinfo {year} {2019}{\natexlab{b}})}\BibitemShut
  {NoStop}%
\bibitem [{\citenamefont {Nosrati}\ \emph
  {et~al.}(2020{\natexlab{a}})\citenamefont {Nosrati}, \citenamefont
  {Castellini}, \citenamefont {Compagno},\ and\ \citenamefont
  {Lo~Franco}}]{saro2020npj}%
  \BibitemOpen
  \bibfield  {author} {\bibinfo {author} {\bibfnamefont {F.}~\bibnamefont
  {Nosrati}}, \bibinfo {author} {\bibfnamefont {A.}~\bibnamefont {Castellini}},
  \bibinfo {author} {\bibfnamefont {G.}~\bibnamefont {Compagno}},\ and\
  \bibinfo {author} {\bibfnamefont {R.}~\bibnamefont {Lo~Franco}},\ }\href
  {https://doi.org/10.1038/s41534-020-0271-7} {\bibfield  {journal} {\bibinfo
  {journal} {npj Quantum Inf.}\ }\textbf {\bibinfo {volume} {6}},\ \bibinfo
  {pages} {1} (\bibinfo {year} {2020}{\natexlab{a}})}\BibitemShut {NoStop}%
\bibitem [{\citenamefont {Nosrati}\ \emph
  {et~al.}(2020{\natexlab{b}})\citenamefont {Nosrati}, \citenamefont
  {Castellini}, \citenamefont {Compagno},\ and\ \citenamefont
  {Lo~Franco}}]{NosratiPRA}%
  \BibitemOpen
  \bibfield  {author} {\bibinfo {author} {\bibfnamefont {F.}~\bibnamefont
  {Nosrati}}, \bibinfo {author} {\bibfnamefont {A.}~\bibnamefont {Castellini}},
  \bibinfo {author} {\bibfnamefont {G.}~\bibnamefont {Compagno}},\ and\
  \bibinfo {author} {\bibfnamefont {R.}~\bibnamefont {Lo~Franco}},\ }\href
  {https://doi.org/10.1103/PhysRevA.102.062429} {\bibfield  {journal} {\bibinfo
   {journal} {Phys. Rev. A}\ }\textbf {\bibinfo {volume} {102}},\ \bibinfo
  {pages} {062429} (\bibinfo {year} {2020}{\natexlab{b}})}\BibitemShut
  {NoStop}%
\bibitem [{\citenamefont {Perez-Leija}\ \emph {et~al.}(2018)\citenamefont
  {Perez-Leija}, \citenamefont {Guzm{\'a}n-Silva}, \citenamefont
  {Le{\'o}n-Montiel}, \citenamefont {Gr{\"a}fe}, \citenamefont {Heinrich},
  \citenamefont {Moya-Cessa}, \citenamefont {Busch},\ and\ \citenamefont
  {Szameit}}]{alexander2018}%
  \BibitemOpen
  \bibfield  {author} {\bibinfo {author} {\bibfnamefont {A.}~\bibnamefont
  {Perez-Leija}}, \bibinfo {author} {\bibfnamefont {D.}~\bibnamefont
  {Guzm{\'a}n-Silva}}, \bibinfo {author} {\bibfnamefont {R.~d.~J.}\
  \bibnamefont {Le{\'o}n-Montiel}}, \bibinfo {author} {\bibfnamefont
  {M.}~\bibnamefont {Gr{\"a}fe}}, \bibinfo {author} {\bibfnamefont
  {M.}~\bibnamefont {Heinrich}}, \bibinfo {author} {\bibfnamefont
  {H.}~\bibnamefont {Moya-Cessa}}, \bibinfo {author} {\bibfnamefont
  {K.}~\bibnamefont {Busch}},\ and\ \bibinfo {author} {\bibfnamefont
  {A.}~\bibnamefont {Szameit}},\ }\href
  {https://doi.org/10.1038/s41534-018-0094-y} {\bibfield  {journal} {\bibinfo
  {journal} {npj Quant. Inf.}\ }\textbf {\bibinfo {volume} {4}},\ \bibinfo
  {pages} {45} (\bibinfo {year} {2018})}\BibitemShut {NoStop}%
\bibitem [{\citenamefont {Sun}\ \emph {et~al.}(2020)\citenamefont {Sun},
  \citenamefont {Wang}, \citenamefont {Liu}, \citenamefont {Xu}, \citenamefont
  {Xu}, \citenamefont {Li}, \citenamefont {Guo}, \citenamefont {Castellini},
  \citenamefont {Nosrati}, \citenamefont {Compagno} \emph {et~al.}}]{sun2020}%
  \BibitemOpen
  \bibfield  {author} {\bibinfo {author} {\bibfnamefont {K.}~\bibnamefont
  {Sun}}, \bibinfo {author} {\bibfnamefont {Y.}~\bibnamefont {Wang}}, \bibinfo
  {author} {\bibfnamefont {Z.-H.}\ \bibnamefont {Liu}}, \bibinfo {author}
  {\bibfnamefont {X.-Y.}\ \bibnamefont {Xu}}, \bibinfo {author} {\bibfnamefont
  {J.-S.}\ \bibnamefont {Xu}}, \bibinfo {author} {\bibfnamefont {C.-F.}\
  \bibnamefont {Li}}, \bibinfo {author} {\bibfnamefont {G.-C.}\ \bibnamefont
  {Guo}}, \bibinfo {author} {\bibfnamefont {A.}~\bibnamefont {Castellini}},
  \bibinfo {author} {\bibfnamefont {F.}~\bibnamefont {Nosrati}}, \bibinfo
  {author} {\bibfnamefont {G.}~\bibnamefont {Compagno}}, \emph {et~al.},\
  }\href {https://doi.org/10.1364/OL.401735} {\bibfield  {journal} {\bibinfo
  {journal} {Opt. Lett.}\ }\textbf {\bibinfo {volume} {45}},\ \bibinfo {pages}
  {6410} (\bibinfo {year} {2020})}\BibitemShut {NoStop}%
\bibitem [{\citenamefont {Napoli}\ \emph {et~al.}(2016)\citenamefont {Napoli},
  \citenamefont {Bromley}, \citenamefont {Cianciaruso}, \citenamefont {Piani},
  \citenamefont {Johnston},\ and\ \citenamefont {Adesso}}]{napoli2016}%
  \BibitemOpen
  \bibfield  {author} {\bibinfo {author} {\bibfnamefont {C.}~\bibnamefont
  {Napoli}}, \bibinfo {author} {\bibfnamefont {T.~R.}\ \bibnamefont {Bromley}},
  \bibinfo {author} {\bibfnamefont {M.}~\bibnamefont {Cianciaruso}}, \bibinfo
  {author} {\bibfnamefont {M.}~\bibnamefont {Piani}}, \bibinfo {author}
  {\bibfnamefont {N.}~\bibnamefont {Johnston}},\ and\ \bibinfo {author}
  {\bibfnamefont {G.}~\bibnamefont {Adesso}},\ }\href
  {https://doi.org/10.1103/PhysRevLett.116.150502} {\bibfield  {journal}
  {\bibinfo  {journal} {Phys. Rev. Lett.}\ }\textbf {\bibinfo {volume} {116}},\
  \bibinfo {pages} {150502} (\bibinfo {year} {2016})}\BibitemShut {NoStop}%
\bibitem [{\citenamefont {Ringbauer}\ \emph {et~al.}(2018)\citenamefont
  {Ringbauer}, \citenamefont {Bromley}, \citenamefont {Cianciaruso},
  \citenamefont {Lami}, \citenamefont {Lau}, \citenamefont {Adesso},
  \citenamefont {White}, \citenamefont {Fedrizzi},\ and\ \citenamefont
  {Piani}}]{ringbauer2018}%
  \BibitemOpen
  \bibfield  {author} {\bibinfo {author} {\bibfnamefont {M.}~\bibnamefont
  {Ringbauer}}, \bibinfo {author} {\bibfnamefont {T.~R.}\ \bibnamefont
  {Bromley}}, \bibinfo {author} {\bibfnamefont {M.}~\bibnamefont
  {Cianciaruso}}, \bibinfo {author} {\bibfnamefont {L.}~\bibnamefont {Lami}},
  \bibinfo {author} {\bibfnamefont {W.~S.}\ \bibnamefont {Lau}}, \bibinfo
  {author} {\bibfnamefont {G.}~\bibnamefont {Adesso}}, \bibinfo {author}
  {\bibfnamefont {A.~G.}\ \bibnamefont {White}}, \bibinfo {author}
  {\bibfnamefont {A.}~\bibnamefont {Fedrizzi}},\ and\ \bibinfo {author}
  {\bibfnamefont {M.}~\bibnamefont {Piani}},\ }\href
  {https://doi.org/10.1103/PhysRevX.8.041007} {\bibfield  {journal} {\bibinfo
  {journal} {Phys. Rev. X}\ }\textbf {\bibinfo {volume} {8}},\ \bibinfo {pages}
  {041007} (\bibinfo {year} {2018})}\BibitemShut {NoStop}%
\bibitem [{\citenamefont {Lo~Franco}\ and\ \citenamefont
  {Compagno}(2016)}]{saro2016}%
  \BibitemOpen
  \bibfield  {author} {\bibinfo {author} {\bibfnamefont {R.}~\bibnamefont
  {Lo~Franco}}\ and\ \bibinfo {author} {\bibfnamefont {G.}~\bibnamefont
  {Compagno}},\ }\href {https://doi.org/10.1038/srep20603} {\bibfield
  {journal} {\bibinfo  {journal} {Sci. Rep.}\ }\textbf {\bibinfo {volume}
  {6}},\ \bibinfo {pages} {20603} (\bibinfo {year} {2016})}\BibitemShut
  {NoStop}%
\bibitem [{\citenamefont {Barros}\ \emph {et~al.}(2020)\citenamefont {Barros},
  \citenamefont {Chin}, \citenamefont {Pramanik}, \citenamefont {Lim},
  \citenamefont {Cho}, \citenamefont {Huh},\ and\ \citenamefont
  {Kim}}]{Barros:20}%
  \BibitemOpen
  \bibfield  {author} {\bibinfo {author} {\bibfnamefont {M.~R.}\ \bibnamefont
  {Barros}}, \bibinfo {author} {\bibfnamefont {S.}~\bibnamefont {Chin}},
  \bibinfo {author} {\bibfnamefont {T.}~\bibnamefont {Pramanik}}, \bibinfo
  {author} {\bibfnamefont {H.-T.}\ \bibnamefont {Lim}}, \bibinfo {author}
  {\bibfnamefont {Y.-W.}\ \bibnamefont {Cho}}, \bibinfo {author} {\bibfnamefont
  {J.}~\bibnamefont {Huh}},\ and\ \bibinfo {author} {\bibfnamefont {Y.-S.}\
  \bibnamefont {Kim}},\ }\href {https://doi.org/10.1364/OE.410361} {\bibfield
  {journal} {\bibinfo  {journal} {Opt. Express}\ }\textbf {\bibinfo {volume}
  {28}},\ \bibinfo {pages} {38083} (\bibinfo {year} {2020})}\BibitemShut
  {NoStop}%
\bibitem [{\citenamefont {Takeuchi}(2001)}]{type2}%
  \BibitemOpen
  \bibfield  {author} {\bibinfo {author} {\bibfnamefont {S.}~\bibnamefont
  {Takeuchi}},\ }\href {https://doi.org/10.1364/OL.26.000843} {\bibfield
  {journal} {\bibinfo  {journal} {Opt. Lett.}\ }\textbf {\bibinfo {volume}
  {26}},\ \bibinfo {pages} {843} (\bibinfo {year} {2001})}\BibitemShut
  {NoStop}%
\bibitem [{\citenamefont {Baumgratz}\ \emph {et~al.}(2014)\citenamefont
  {Baumgratz}, \citenamefont {Cramer},\ and\ \citenamefont
  {Plenio}}]{plenio2014}%
  \BibitemOpen
  \bibfield  {author} {\bibinfo {author} {\bibfnamefont {T.}~\bibnamefont
  {Baumgratz}}, \bibinfo {author} {\bibfnamefont {M.}~\bibnamefont {Cramer}},\
  and\ \bibinfo {author} {\bibfnamefont {M.~B.}\ \bibnamefont {Plenio}},\
  }\href {https://doi.org/10.1103/PhysRevLett.113.140401} {\bibfield  {journal}
  {\bibinfo  {journal} {Phys. Rev. Lett.}\ }\textbf {\bibinfo {volume} {113}},\
  \bibinfo {pages} {140401} (\bibinfo {year} {2014})}\BibitemShut {NoStop}%
\bibitem [{\citenamefont {Piani}\ \emph {et~al.}(2016)\citenamefont {Piani},
  \citenamefont {Cianciaruso}, \citenamefont {Bromley}, \citenamefont {Napoli},
  \citenamefont {Johnston},\ and\ \citenamefont {Adesso}}]{piani2016}%
  \BibitemOpen
  \bibfield  {author} {\bibinfo {author} {\bibfnamefont {M.}~\bibnamefont
  {Piani}}, \bibinfo {author} {\bibfnamefont {M.}~\bibnamefont {Cianciaruso}},
  \bibinfo {author} {\bibfnamefont {T.~R.}\ \bibnamefont {Bromley}}, \bibinfo
  {author} {\bibfnamefont {C.}~\bibnamefont {Napoli}}, \bibinfo {author}
  {\bibfnamefont {N.}~\bibnamefont {Johnston}},\ and\ \bibinfo {author}
  {\bibfnamefont {G.}~\bibnamefont {Adesso}},\ }\href
  {https://doi.org/10.1103/PhysRevA.93.042107} {\bibfield  {journal} {\bibinfo
  {journal} {Phys. Rev. A}\ }\textbf {\bibinfo {volume} {93}},\ \bibinfo
  {pages} {042107} (\bibinfo {year} {2016})}\BibitemShut {NoStop}%
\bibitem [{\citenamefont {Marvian}\ and\ \citenamefont
  {Spekkens}(2016)}]{spekkens2016}%
  \BibitemOpen
  \bibfield  {author} {\bibinfo {author} {\bibfnamefont {I.}~\bibnamefont
  {Marvian}}\ and\ \bibinfo {author} {\bibfnamefont {R.~W.}\ \bibnamefont
  {Spekkens}},\ }\href {https://doi.org/10.1103/PhysRevA.94.052324} {\bibfield
  {journal} {\bibinfo  {journal} {Physical Review A}\ }\textbf {\bibinfo
  {volume} {94}},\ \bibinfo {pages} {052324} (\bibinfo {year}
  {2016})}\BibitemShut {NoStop}%
\bibitem [{\citenamefont {Wang}\ \emph {et~al.}(2017)\citenamefont {Wang},
  \citenamefont {Tang}, \citenamefont {Wei}, \citenamefont {Yu}, \citenamefont
  {Ke}, \citenamefont {Xu}, \citenamefont {Li},\ and\ \citenamefont
  {Guo}}]{wang2017}%
  \BibitemOpen
  \bibfield  {author} {\bibinfo {author} {\bibfnamefont {Y.-T.}\ \bibnamefont
  {Wang}}, \bibinfo {author} {\bibfnamefont {J.-S.}\ \bibnamefont {Tang}},
  \bibinfo {author} {\bibfnamefont {Z.-Y.}\ \bibnamefont {Wei}}, \bibinfo
  {author} {\bibfnamefont {S.}~\bibnamefont {Yu}}, \bibinfo {author}
  {\bibfnamefont {Z.-J.}\ \bibnamefont {Ke}}, \bibinfo {author} {\bibfnamefont
  {X.-Y.}\ \bibnamefont {Xu}}, \bibinfo {author} {\bibfnamefont {C.-F.}\
  \bibnamefont {Li}},\ and\ \bibinfo {author} {\bibfnamefont {G.-C.}\
  \bibnamefont {Guo}},\ }\href {https://doi.org/10.1103/PhysRevLett.118.020403}
  {\bibfield  {journal} {\bibinfo  {journal} {Phys. Rev. Lett.}\ }\textbf
  {\bibinfo {volume} {118}},\ \bibinfo {pages} {020403} (\bibinfo {year}
  {2017})}\BibitemShut {NoStop}%
\bibitem [{\citenamefont {Barnett}\ and\ \citenamefont
  {Croke}(2009)}]{croke2009}%
  \BibitemOpen
  \bibfield  {author} {\bibinfo {author} {\bibfnamefont {S.~M.}\ \bibnamefont
  {Barnett}}\ and\ \bibinfo {author} {\bibfnamefont {S.}~\bibnamefont
  {Croke}},\ }\href {https://doi.org/10.1364/AOP.1.000238} {\bibfield
  {journal} {\bibinfo  {journal} {Adv. Opt. Photonics}\ }\textbf {\bibinfo
  {volume} {1}},\ \bibinfo {pages} {238} (\bibinfo {year} {2009})}\BibitemShut
  {NoStop}%
\bibitem [{\citenamefont {Mosley}\ \emph {et~al.}(2006)\citenamefont {Mosley},
  \citenamefont {Croke}, \citenamefont {Walmsley},\ and\ \citenamefont
  {Barnett}}]{mosley2006}%
  \BibitemOpen
  \bibfield  {author} {\bibinfo {author} {\bibfnamefont {P.~J.}\ \bibnamefont
  {Mosley}}, \bibinfo {author} {\bibfnamefont {S.}~\bibnamefont {Croke}},
  \bibinfo {author} {\bibfnamefont {I.~A.}\ \bibnamefont {Walmsley}},\ and\
  \bibinfo {author} {\bibfnamefont {S.~M.}\ \bibnamefont {Barnett}},\ }\href
  {https://doi.org/10.1103/PhysRevLett.97.193601} {\bibfield  {journal}
  {\bibinfo  {journal} {Phys. Rev. Lett.}\ }\textbf {\bibinfo {volume} {97}},\
  \bibinfo {pages} {193601} (\bibinfo {year} {2006})}\BibitemShut {NoStop}%
\bibitem [{\citenamefont {Helstrom}(1969)}]{Helstrom}%
  \BibitemOpen
  \bibfield  {author} {\bibinfo {author} {\bibfnamefont {C.~W.}\ \bibnamefont
  {Helstrom}},\ }\href@noop {} {\bibfield  {journal} {\bibinfo  {journal} {J.
  Stat. Phys.}\ }\textbf {\bibinfo {volume} {1}},\ \bibinfo {pages} {231}
  (\bibinfo {year} {1969})}\BibitemShut {NoStop}%
\bibitem [{\citenamefont {Holevo}(1978)}]{Holevo}%
  \BibitemOpen
  \bibfield  {author} {\bibinfo {author} {\bibfnamefont {A.~S.}\ \bibnamefont
  {Holevo}},\ }\href@noop {} {\bibfield  {journal} {\bibinfo  {journal} {Proc.
  Steklov Inst. Math.}\ }\textbf {\bibinfo {volume} {124}},\ \bibinfo {pages}
  {1} (\bibinfo {year} {1978})}\BibitemShut {NoStop}%
\bibitem [{\citenamefont {Barnett}\ and\ \citenamefont
  {Riis}(1997)}]{barnett1997}%
  \BibitemOpen
  \bibfield  {author} {\bibinfo {author} {\bibfnamefont {S.~M.}\ \bibnamefont
  {Barnett}}\ and\ \bibinfo {author} {\bibfnamefont {E.}~\bibnamefont {Riis}},\
  }\href {https://doi.org/10.1080/09500349708230718} {\bibfield  {journal}
  {\bibinfo  {journal} {J. Mod. Opt.}\ }\textbf {\bibinfo {volume} {44}},\
  \bibinfo {pages} {1061} (\bibinfo {year} {1997})}\BibitemShut {NoStop}%
\bibitem [{\citenamefont {Mohseni}\ \emph {et~al.}(2004)\citenamefont
  {Mohseni}, \citenamefont {Steinberg},\ and\ \citenamefont
  {Bergou}}]{mohseni2004}%
  \BibitemOpen
  \bibfield  {author} {\bibinfo {author} {\bibfnamefont {M.}~\bibnamefont
  {Mohseni}}, \bibinfo {author} {\bibfnamefont {A.~M.}\ \bibnamefont
  {Steinberg}},\ and\ \bibinfo {author} {\bibfnamefont {J.~A.}\ \bibnamefont
  {Bergou}},\ }\href {https://doi.org/10.1103/PhysRevLett.93.200403} {\bibfield
   {journal} {\bibinfo  {journal} {Phys. Rev. Lett.}\ }\textbf {\bibinfo
  {volume} {93}},\ \bibinfo {pages} {200403} (\bibinfo {year}
  {2004})}\BibitemShut {NoStop}%
\bibitem [{\citenamefont {Sciara}\ \emph {et~al.}(2017)\citenamefont {Sciara},
  \citenamefont {Lo~Franco},\ and\ \citenamefont {Compagno}}]{SciaraSciRep}%
  \BibitemOpen
  \bibfield  {author} {\bibinfo {author} {\bibfnamefont {S.}~\bibnamefont
  {Sciara}}, \bibinfo {author} {\bibfnamefont {R.}~\bibnamefont {Lo~Franco}},\
  and\ \bibinfo {author} {\bibfnamefont {G.}~\bibnamefont {Compagno}},\ }\href
  {https://doi.org/10.1038/srep44675} {\bibfield  {journal} {\bibinfo
  {journal} {Sci. Rep.}\ }\textbf {\bibinfo {volume} {7}},\ \bibinfo {pages}
  {44675} (\bibinfo {year} {2017})}\BibitemShut {NoStop}%
\end{thebibliography}

\providecommand{\noopsort}[1]{}\providecommand{\singleletter}[1]{#1}%
%

\end{document}